\documentclass[final,5p,times,twocolumn]{elsarticle}

\usepackage[T1]{fontenc}
\usepackage[utf8]{inputenc}

\usepackage{amsmath}
\usepackage{amsfonts}
\usepackage{amssymb}
\usepackage{amsxtra}
\usepackage{array}
\usepackage{color}
\usepackage{dcolumn}
\usepackage{graphicx}
\usepackage{hepunits}
\usepackage{xspace}
\usepackage{CJKutf8}

\definecolor{purple}{rgb}{0.5,0,0.5}
\definecolor{blue}{rgb}{0.0,0,0.9}
\definecolor{prdblue}{rgb}{0.133,0.118,0.498}
\usepackage[colorlinks=true, pdfstartview=FitV, linkcolor=prdblue, citecolor= prdblue, urlcolor=prdblue]{hyperref}

\usepackage[mathscr,scaled=1.15]{urwchancal}
\DeclareFontFamily{OT1}{pzc}{}
\DeclareFontShape{OT1}{pzc}{m}{it}%
{<-> s * [1.15] pzcmi7t}{}
\DeclareMathAlphabet{\mathpzc}{OT1}{pzc}{m}{it}

\biboptions{sort&compress}

\journal{Physics Letters B}

\makeatletter

\setbox0\hbox{$\xdef\scriptratio{\strip@pt\dimexpr
    \numexpr(\sf@size*65536)/\f@size sp}$}

\newcommand{\scriptveryshortarrow}[1][3pt]{{%
    \hbox{\rule[\scriptratio\dimexpr\fontdimen22\textfont2-.2pt\relax]
               {\scriptratio\dimexpr#1\relax}{\scriptratio\dimexpr.4pt\relax}}%
   \mkern-4mu\hbox{\let\f@size\sf@size\usefont{U}{lasy}{m}{n}\symbol{41}}}}

\makeatother

\begin{document}
\begin{CJK}{UTF8}{song}

\begin{frontmatter}

\title{$\,$\\[-7ex]\hspace*{\fill}{\normalsize{\sf\emph{Preprint no}. NJU-INP 089/24}}\\[1ex]
$J/\psi$ photoproduction: threshold to very high energy}

\author{Lin Tang
       $^{\href{https://orcid.org/0009-0001-2324-9963}{\textcolor[rgb]{0.00,1.00,0.00}{\sf ID}}}$}

\author{Yi-Xuan Yang
       $^{\href{https://orcid.org/0009-0006-6946-6679}{\textcolor[rgb]{0.00,1.00,0.00}{\sf ID}}}$}

\author{Zhu-Fang~Cui
       $^{\href{https://orcid.org/0000-0003-3890-0242}{\textcolor[rgb]{0.00,1.00,0.00}{\sf ID}}}$}

\author{Craig D. Roberts%
       $^{\href{https://orcid.org/0000-0002-2937-1361}{\textcolor[rgb]{0.00,1.00,0.00}{\sf ID}}}$}

%
\address{
School of Physics, Nanjing University, Nanjing, Jiangsu 210093, China\\
%
Institute for Nonperturbative Physics, Nanjing University, Nanjing, Jiangsu 210093, China\\[1ex]
%
\href{mailto:phycui@nju.edu.cn}{phycui@nju.edu.cn} (Z.-F. Cui);
\href{mailto:cdroberts@nju.edu.cn}{cdroberts@nju.edu.cn} (C. D. Roberts)
\\[1ex]
Date: 2024 July 19\\[-6ex]
}

\begin{abstract}
A reaction model for $\gamma + p \to J/\psi + p$ photoproduction, which exposes the $c \bar c$ content of the photon in making the transition $\gamma\to c\bar c + \mathbb P \to J/\psi$ and couples the intermediate $c \bar c$ system to the proton's valence quarks via Pomeron ($\mathbb P $) exchange, is used to deliver a description of available data, \emph{viz}.\ both differential and total cross sections from near threshold, where data has newly been acquired, to invariant mass $W \approx 300\,$GeV.  The study suggests that it is premature to link existing $\gamma + p \to J/\psi + p$ data with, for instance, in-proton gluon distributions, the quantum chromodynamics trace anomaly, or pentaquark production.  Further developments in reaction theory and higher precision data are necessary before the validity of any such connections can be assessed.
\end{abstract}


\end{frontmatter}
\end{CJK}


\section{Introduction}
%
Consider a $V = Q+\bar Q$ vector meson, with valence degrees-of-freedom $Q=c$, $b$, \emph{i.e}., a heavy vector meson.  Such systems have no valence components in common with the proton (nucleon).  Consequently, it is widely held \cite{Krein:2017usp, Lee:2022ymp} that the dominant mechanisms underlying photoproduction of these systems from the proton must involve some manifestations of gluon physics within the target proton and, perhaps, the incipient vector meson, and/or the exchange of (perhaps infinitely many, correlated) gluons between the partonic constituents of each.

If one presupposes that vector meson dominance (VMD) can be used to reinterpret $\gamma + p \to V + p$ in terms of $V + p \to V + p$, then such photoproduction reactions near-threshold seem \cite{Kharzeev:1995ij} to provide experimental access to the in-proton expectation value of the trace anomaly in quantum chromodynamics (QCD); hence, insights into the character of emergent hadron mass \cite{Roberts:2016vyn, Krein:2020yor, Roberts:2021nhw, Binosi:2022djx, Ding:2022ows, Ferreira:2023fva, Salme:2022eoy, Carman:2023zke}.  This possibility has served as a motivation for new high-energy, high-luminosity accelerator facilities \cite{Chen:2020ijn, Anderle:2021wcy, AbdulKhalek:2021gbh}.  Assuming VMD to be a valid reaction model, then there is also a potential connection between threshold $\gamma + p \to V + p$ reactions and photoproduction of pentaquark states, like those reported in Ref.\,\cite[LHCb]{LHCb:2015yax}.
However, such possibilities now appear remote given that modern analyses \cite{Du:2020bqj, Xu:2021mju, Sun:2021pyw} have shown that VMD cannot be used in these ways.

Nevertheless, even eschewing VMD, some analyses advocate that it may still be possible to interpret near-threshold $\gamma + p \to V + p$ reactions using generalised parton distributions (GPDs) \cite{Guo:2023pqw} or holographic models \cite{Mamo:2022eui} and therewith gain access to in-proton gluon gravitational form factors.
Yet, even here, there are counter arguments \cite{Du:2020bqj}, which continue to be developed.
For instance, considering a reaction mechanism based on a phenomenological quark + nucleon potential and Pomeron ($\mathbb P$) exchange, it has been suggested that $J/\psi \, N$ final state interactions (FSIs) dominate the cross-section near-threshold \cite{Sakinah:2024cza}.  If correct, then this obscures any connection with in-proton gluon distributions and/or gluon gravitational form factors.

Evidently, given the desire to interpret $\gamma + p \to V + p$ photoproduction data in terms of gluon physics, an elucidation of the underlying reaction mechanism is crucial.
This point was argued, \emph{e.g}., in Refs.\,\cite[GlueX]{GlueX:2019mkq, GlueX:2023pev}, which collected precise cross-section data on $\gamma + p \to J/\psi + p$ from threshold, \emph{viz}.\ invariant mass $W_{\rm th}=m_{J/\psi}+m_p = 4.04\,$GeV, to $W=4.72\,$GeV and over a large kinematic range of momentum transfer, $t$.  (\emph{N.B}.\ For a target proton at rest, this $W$ coverage corresponds to a photon energy range $\nu/{\rm GeV} = 8.2 - 11.4$.)
Near-threshold data are also provided in Ref.\,\cite[$J/\psi$-007]{Duran:2022xag}; and data (far) beyond threshold have long existed \cite{Camerini:1975cy, Gittelman:1975ix, Shambroom:1982qj, ZEUS:1995kab, H1:1996gwv, Amarian:1999pi, H1:2000kis, ZEUS:2002wfj}, reaching to $W=300\,$GeV.

At issue today is the challenge of delivering a unified description of all available data.  Hitherto, reaction models which describe the high-$W$ data typically disagree with near-threshold data, and vice-versa \cite{Lee:2022ymp}.  Herein, we approach this challenge by expanding upon an earlier analysis \cite{Pichowsky:1996tn}, which resolved the $\gamma\to c\bar c + \mathbb P \to J/\psi$ transition loop in terms of dressed-quark/antiquark degrees-of-freedom and expressed the low momentum transfer interaction between these loop constituents and the proton via Pomeron exchange. 

\begin{figure}[t]
\centerline{%
\includegraphics[clip, width=0.3\textwidth]{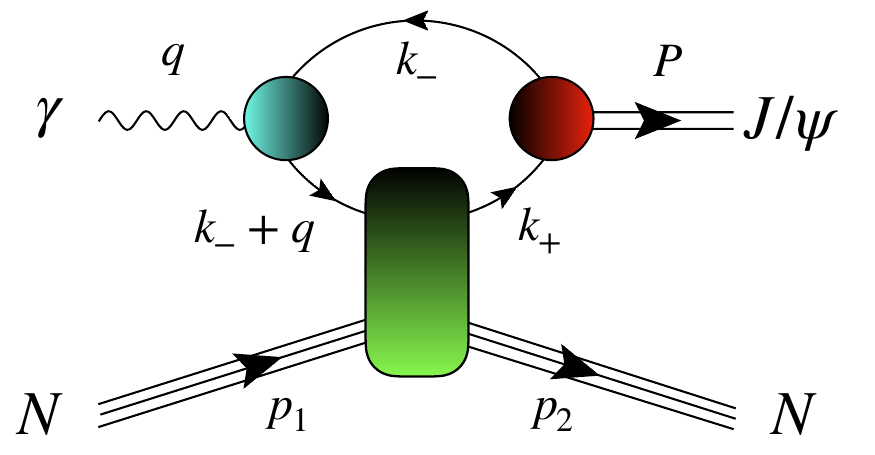}}
\caption{\label{FigPomeron}
Reaction model for $\gamma + p \to J/\psi + p$.
The $c\bar c$ component of the dressed-photon is probed by Pomeron exchange with the proton, producing an on-shell $J/\psi$ meson.
Referring to Eq.\,\eqref{EqtqP}:
the green rectangle describes Pomeron exchange, $\Gamma^{\mathbb P}$, including its couplings to the proton and quark -- see Sect.\,\ref{SecPomeron};
the solid black curve is $S_c$, the dressed $c$ quark propagator;
the shaded blue circle is $\Gamma_\mu^\gamma$, the dressed-$\gamma c \bar c$ vertex;
and the shaded red circle is $\Gamma_\nu$, $J/\psi$-meson Bethe-Salpeter amplitude -- see Sect.\,\ref{SecPMV}.
Kinematics: $s=W^2 = -(p_2^2+ P^2)$; $t=-(p_2-p_1)^2$.
%
(Image courtesy of D.~Binosi.)
}
\end{figure}

\section{Reaction model}
Our model for the $J/\psi$ photoproduction reaction is sketched in Fig.\,\ref{FigPomeron}.
The image corresponds to the following matrix element \cite{Pichowsky:1996tn, Donnachie:1984xq, Donnachie:1987pu}:
\begin{align}
{\cal I}_\mu(W,t) & = \langle J(P;\lambda) p(p_2) | \bar c \gamma_\mu c| p(p_1)\rangle  \nonumber \\
& =
2 t_{\mu \alpha \nu}(q,P) \epsilon_\nu^\lambda(P) \bar u(p_2) \tilde G_\alpha(w^2,t) u(p_1)\,,
\end{align}
where
$\epsilon_\nu^\lambda(P)$ is the $J/\psi$ polarisation vector,
$u(p_1)$ and $\bar u(p_2)$ are spinors for the incoming and outgoing proton,
$\tilde G_\alpha(w^2,t) $ represents the proton-Pomeron-quark interaction;
and the factor of ``2'' accounts for the equivalence between striking the upper and lower valence quark lines.

Capitalising on the fact that the explicitly exposed quark loop integration in Fig.\,\ref{FigPomeron} receives most support in the neighbourhood $k^2 \simeq 0$, \emph{i.e}., on the domain of near-zero relative momentum within the final state meson,
one may write $w^2=-(q-P/2+p_1)^2$.  Then the essentially dynamical component of the reaction model, \emph{viz}.\ the $\gamma\to c\bar c + \mathbb P \to J/\psi$ transition matrix element, can be written as follows:
\begin{align}
t_{\mu \alpha \nu}(q,P) & = \tfrac{2}{3} e N_c
{\rm tr}_D \int \frac{d^4 k}{(2\pi)^4}
S_c(k_-) \Gamma_\mu^\gamma(k_-,k_-+q) \nonumber \\
& \quad \times S_c(k_-+q) \Gamma^{\mathbb P}_\alpha(k_-+q,k_+) S_c(k_+) \Gamma_\nu(k_+,k_-)\,,
\label{EqtqP}
\end{align}
where $N_c=3$,
the fine-structure constant $\alpha = e^2/[4\pi]$,
the trace is over spinor indices,
$k_\pm = k \pm P/2$,
$S_c$ is the dressed $c$ quark propagator,
$\Gamma_\mu^\gamma$ is the associated dressed photon-quark vertex,
$ \Gamma^{\mathbb P}_\alpha$ is the Pomeron-quark vertex,
and
$\Gamma_\nu$ is the Bethe-Salpeter amplitude for an on-shell $J/\psi$ meson.

Working in the $J/\psi + p$ centre-of-momentum frame, one readily arrives at the differential cross-section for photoproduction ($q^2=0$) \cite{Pichowsky:1996tn}:
\begin{align}
\frac{d\sigma}{d\Omega} & =
\frac{1}{4\pi^2} \frac{m_p}{4 W} \frac{|\vec{P}|}{k_W}
\sum_{\rm proton\;spin} (|{\cal I}_1|^2 + |{\cal I}_2|^2)\,,
\end{align}
where
$q=(\vec q, i |\vec q| )$,
$P = (\vec{P}, i E_J )$, $E_J^2 =  m_{J/\psi}^2+|\vec P|^2$,
$k_W = (W^2 - m_p^2)/[2 m_p]$,
$|t|_{\rm min} = 2 |\vec{q}| (|\vec{P}|^2+m_{J/\psi}^2)^{1/2}-m_{J/\psi}^2 - 2 |\vec{q}| |\vec{P}|$,
and
$d\Omega$ is the differential solid-angle element defined by $\angle \vec{q}  \vec P$.
Naturally,
\begin{align}
\frac{d\sigma}{d(-t)} & = \frac{\pi}{|\vec P| |\vec q|} \frac{d\sigma}{d\Omega}\,.
\end{align}
For photoproduction, the longitudinal cross-section vanishes.

\section{Pomeron Model}
\label{SecPomeron}
The reaction model involves two Pomeron elements.

The Pomeron-proton content of the coupling is \cite{Pichowsky:1996tn, Donnachie:1984xq, Donnachie:1987pu}:
\begin{equation}
\label{EqGP}
\tilde G_\alpha(s,t) = \gamma_\alpha F_0(t)  3 \beta_\ell G_{\mathbb P}(s,t) \,,
\end{equation}
where
$F_0(t)$ is the empirical nucleon isoscalar elastic electromagnetic form factor, parametrised efficaciously as follows on the spacelike low-$|t|$ domain relevant herein:
\begin{equation}
F_0(t) = \frac{1-2.8 t/[4m_p^2]}{1-t/[4m_p^2]} \frac{1}{(1-t/t_0)^2}\,,
\end{equation}
$t_0 = 0.71\,$GeV$^2$;
``$3$'' counts the number of light valence degrees-of-freedom in the proton;
and $\beta_\ell$ is the strength of the associated Pomeron + light-quark coupling, which may be determined, \emph{e.g}., via a fit to the  differential cross-section for $\rho^0$-meson photoproduction \cite{Pichowsky:1996tn}.

The remaining element in Eq.\,\eqref{EqGP} is the object typically called the Pomeron propagator:
\begin{equation}
G_{\mathbb P}(s,t) = \left[\frac{s}{s_0}\right]^{\alpha_{\mathbb P}(t) - 1}
\exp\left[ - i \tfrac{\pi}{2} (\alpha_{\mathbb P}(t) - 1) \right]\,,
\end{equation}
with
$\alpha_{\mathbb P}(t) = \alpha_0 + \alpha_1 t$,
$s_0 = 1/\alpha_1$,
and $\alpha_{0,1}$ fitted to selected small-$|t|$, large-$W$ meson photoproduction data within the context of Regge phenomenology -- see, \emph{e.g}., Refs.\,\cite{Collins:1977jy, Irving:1977ea}.
These parameters depend on the final state vector meson, showing differences between light- and heavy-meson final states \cite{Lee:2022ymp}.

It is worth remarking that we describe the proton as a bound state of three dressed light quarks.  The possibility of ``intrinsic'' strange and/or charm ($q + \bar q$, $q=s, c$) components in the proton (light-front) wave function was highlighted in Ref.\,\cite{Brodsky:1980pb}.  This remains a topical issue, as stressed, \emph{e.g}., in Refs.\,\cite{Sufian:2020coz, Olamaei:2023bpi}.  A contemporary phenomenological fit to a selection of available data \cite{Ball:2022qks} may be interpreted as supporting the existence of a small intrinsic charm component in the proton and a corresponding in-proton $c+\bar c$ light-front momentum fraction of roughly $0.6$\% at a typical resolving scale.  We remain circumspect on this point, however, because analyses completed using continuum Schwinger function methods (CSMs) predict a commensurate value of this momentum fraction without recourse to intrinsic charm \cite{Lu:2022cjx, Yu:2024qsd}.  Instead, the fraction is produced entirely via gluon radiation of $c+\bar c$ pairs, the momentum distribution profile of which is sea quark like.

Returning to our reaction model, which exposes the dressed-photon's quark content via the $\gamma \to \bar c c + {\mathbb P} \to J/\psi$ loop in Eq.\,\eqref{EqtqP}, then, relative to the proton coupling, the Pomeron-quark coupling in the loop is much simpler \cite{Pichowsky:1996tn, Donnachie:1984xq, Donnachie:1987pu}: $\Gamma^{\mathbb P}_\alpha(k_-+q,k_+)
= \beta_c \gamma_\alpha$.
(More generally, $\beta_c\to \beta_f$, where $f$ is the flavour of quark in the loop and the final-state meson.)

At this point, there are four Pomeron-related parameters:
$\beta_{\ell,c}$, $\alpha_{0,1}$.
Their determination is discussed in Sect.\,\ref{SecPomParams}.

\section{Photon to meson transition vertex}
\label{SecPMV}
The transition vertex in Eq.\,\eqref{EqtqP} involves three nonperturbative Schwinger functions.  In Ref.\,\cite{Pichowsky:1996tn}, owing principally to limitations imposed by then available computational resources, the Dirac structures of these elements were simplified and the associated scalar functions represented by algebraic parametrisations in order to simplify calculation of the loop integral and arrive at a reasonable representation of $t_{\mu\alpha\nu}$.  We have validated our analysis by reproducing all results therein, including those for $\rho$- and $\phi$-meson production.
(Ref.\,\cite{Pichowsky:1996tn} inadvertently introduced $w^2 =-(q + P/2 + p_1)^2$.  We used this value to reproduce the results therein, but the correct expression for the predictions described below.)

Today, the matrix-valued functions in Eq.\,\eqref{EqtqP} are readily computed using CSMs \cite{Binosi:2022djx, Ding:2022ows, Ferreira:2023fva}, so it is now possible to arrive at a parameter-free prediction for $t_{\mu \alpha \nu}(q,P)$.  The key in such calculations is the quark+antiquark scattering kernel, for which the leading-order (rainbow-ladder, RL) truncation \cite{Munczek:1994zz, Bender:1996bb} is obtained by writing \cite{Maris:1997tm}:
{\allowdisplaybreaks
\label{EqRLInteraction}
\begin{align}
\label{KDinteraction}
\mathscr{K}_{tu}^{rs}(k) & =
\tilde{\mathpzc G}(y)
[i\gamma_\mu\frac{\lambda^{a}}{2} ]_{ts} [i\gamma_\nu\frac{\lambda^{a}}{2} ]_{ru} T_{\mu\nu}(k)\,,
%
\end{align}
$k^2 T_{\mu\nu}(k) = k^2 \delta_{\mu\nu} - k_\mu k_\nu$,  $y=k^2$.  The tensor structure corresponds to Landau gauge, used because it is both a fixed point of the renormalisation group and that gauge for which corrections to RL truncation are least noticeable \cite{Bashir:2009fv}.
In Eq.\,\eqref{EqRLInteraction}, $r,s,t,u$ represent colour, spinor, flavour matrix indices (as necessary).
}

A realistic form of $\tilde{\mathpzc G}(y) $ is explained in Refs.\,\cite{Qin:2011dd, Binosi:2014aea}:
\begin{align}
\label{defcalG}
 \tilde{\mathpzc G}(y) & =
 \frac{8\pi^2}{\omega^4} D e^{-y/\omega^2} + \frac{8\pi^2 \gamma_m \mathcal{F}(y)}{\ln\big[ \tau+(1+y/\Lambda_{\rm QCD}^2)^2 \big]}\,,
\end{align}
$\gamma_m=12/25$, $\tau={\rm e}^2-1$, $\Lambda_{\rm QCD}=0.234\,$GeV, and ${\cal F}(y) = \{1 - \exp(-y/\Lambda_{\mathpzc I}^2)\}/y$, $\Lambda_{\mathpzc I}=1\,$GeV.
%
%
In solving all relevant Dyson-Schwinger equations, we use a mass-independent (chiral-limit) momentum-subtraction renormalisation scheme \cite{Chang:2008ec}. 

Widespread application has established \cite{Ding:2022ows} that interactions in the class typified by Eqs.\,\eqref{EqRLInteraction}, \eqref{defcalG} can serve to unify the properties of many systems.
Importantly, when $\omega D=:\varsigma^3$ is held fixed, results for observables remain practically unchanged under $\omega \to (1\pm 0.2)\omega$ \cite{Qin:2020rad}.
Thus, the interaction in Eq.\,\eqref{defcalG} is determined by just one parameter: $\varsigma^3:=\omega D$.

\begin{table}[t]
\caption{\label{TabStatic}
Parameters defining the interaction in Eqs.\,\eqref{EqRLInteraction}, \eqref{defcalG} for distinct renormalisation point invariant (RPI) quark current-masses along with resulting values for the associated vector meson masses and leptonic decay constants.
Empirical values (in GeV) for comparison, drawn from Ref.\,\cite{Workman:2022ynf}:
$m_\rho = 0.775(1)$, $f_\rho =0.153(1)$;
$m_\phi = 1.019(1)$, $f_\phi =0.168(1)$;
$m_{J/\psi} = 3.097(1)$, $f_\phi =0.294(5)$.
One-loop evolved to $\zeta_2 = 2\,$GeV, the RPI current masses listed below correspond to, respectively, in GeV,
$m_l^{\zeta_2}=0.0048$, $m_s^{\zeta_2}=0.117$, $m_c^{\zeta_2}=1.27$.  These values are commensurate with those obtained via other methods \cite{Workman:2022ynf}.
(Isospin symmetry is assumed throughout.  Dimensioned quantities in GeV.)}
\begin{center}
\begin{tabular*}
{\hsize}
{
l@{\extracolsep{0ptplus1fil}}
|l@{\extracolsep{0ptplus1fil}}
l@{\extracolsep{0ptplus1fil}}
l@{\extracolsep{0ptplus1fil}}
|l@{\extracolsep{0ptplus1fil}}
l@{\extracolsep{0ptplus1fil}}}\hline\hline
          & $\hat m$ & $\varsigma$ & $\omega$ & $m_V$ & $f_V$ \\\hline
$\ell=u=d\ $  & $0.0069$ & $0.8$ & $0.5\ $  & $0.741 $ & $0.145$ \\
$s$ & $0.169\phantom{0}$ & $0.8$ & $0.5\ $  & $1.086$ & $0.183$ \\
$c$ & $1.83\phantom{00}$ & $0.6$ & $0.8\ $ & $3.123$ & $0.278$
 \\\hline\hline
\end{tabular*}
\end{center}
\end{table}

In the $u$, $d$, $s$ quark sector, $\varsigma_q  =0.8\,{\rm GeV}$ delivers a good description of a range of pseudoscalar and vector meson static properties -- see Table~\ref{TabStatic} and, \emph{e.g}., Refs.\,\cite{Chen:2018rwz, Ding:2018xwy, Xu:2019ilh}.
For heavier quarks, on the other hand, it is known that corrections to RL truncation are significantly diminished.  Thus, the interaction parameter should be closer to that used in more sophisticated kernels.  This is discussed further in, \emph{e.g}., Ref.\,\cite[Sect.\,IIB]{Xu:2019ilh}, and explains the $c$ quark values listed in Table~\ref{TabStatic}.

It is now time-consuming but straightforward to evaluate the $\gamma \to \bar c c + {\mathbb P} \to V$ loop in Fig.\,\ref{FigPomeron} for the photoproduction of $\rho$, $\phi$, $J/\psi$ mesons.
Using standard numerical procedures \cite{Maris:1997tm, Krassnigg:2009gd},
one calculates the relevant quark propagator, photon-quark vertex, and vector meson Bethe-Salpeter amplitude; then combines them to produce the integrand;
and subsequently evaluates the integral as a function of $(W,t)$.
Owing to current-conservation and the fact that $P_\nu \Gamma_\nu = 0$, $t_{\mu\alpha\nu}$ involves only $9$ independent scalar functions.  Each one is readily extracted using sensibly chosen projection operators.

In all calculations that follow, $A_1(W,t)$, the scalar function which multiplies the tensor structure $t_{\mu\alpha\nu}^1= q_\alpha [\delta_{\mu\nu} + P_\mu P_\nu/m_{J/\psi}^2]$ in $t_{\mu\alpha\nu}$, provides the overwhelmingly dominant contribution -- see below.
Consequently,
excellent approximations may be obtained by projecting onto and retaining only $t_{\mu\alpha\nu}^1$.  We nevertheless keep all nine terms.  For reference, we record the following simple, reliable approximations ($W_0=4.5\,$GeV):
\begin{subequations}
\label{A1Wt}
\begin{align}
m_{J/\psi} \, A_1(W_0 ,|t|) &
\stackrel{0.5 < |t|/{\rm GeV}^2 < 8}{=} 
\frac{3.76\, -0.0110 |t|}{1+0.0845 |t|} 10^{-3}\,,\\
m_{J/\psi}\,  A_1(W,|t|_{\rm min}) & \stackrel{W>W_{\rm th}}{=}
0.00376 \left(1-\frac{W_{\rm th}}{W}\right)^{0.0519} \nonumber \\
& \quad \times  \frac{0.203 W^2 + 0.0509 W - 1}{0.203 W^2-1}\,.
\end{align}
\end{subequations}

\begin{figure}[t]
\centerline{%
\includegraphics[clip, width=0.46\textwidth]{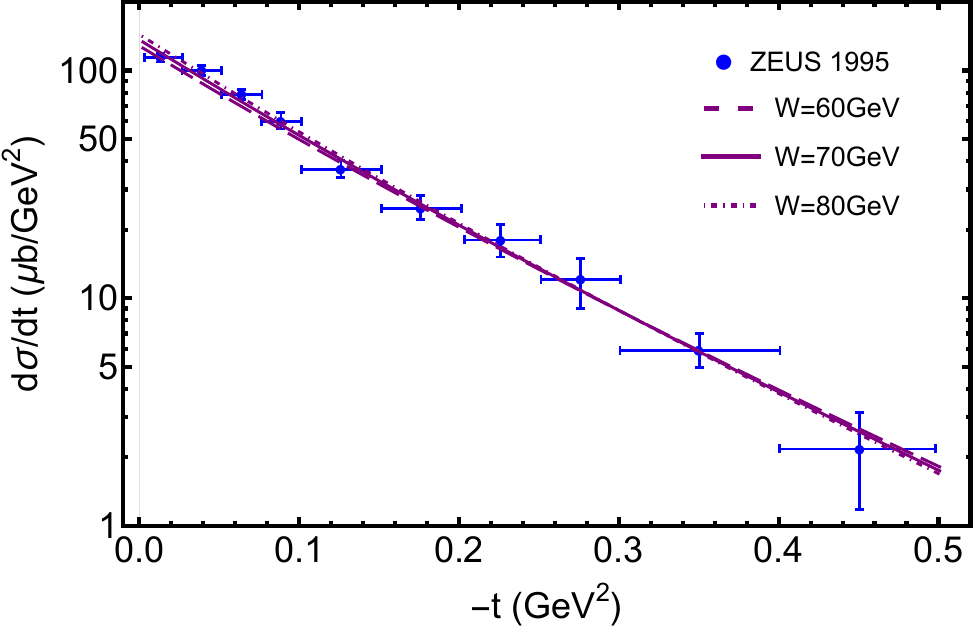}}
\caption{\label{Figrho0}
Differential cross-section for $\gamma p \to \rho^0 p$ at $W\approx 70\,$GeV: the results display little sensitivity to $W$ over the measured range.
All curves: results obtained herein using the reaction model sketched in Fig.\,\ref{FigPomeron} and Pomeron parameters in Table~\ref{TabPom}\,--\,Row~1.
Data from Ref.\,\cite{ZEUS:1995bfs}.
}
\end{figure}

\begin{table}[t]
\caption{\label{TabPom}
Parameters characterising the Pomeron exchange interaction in Fig.\,\ref{FigPomeron}, drawn from analyses in Refs.\,\cite{Donnachie:1987pu, Cudell:1996sh, Pichowsky:1996tn, ZEUS:2002wfj, Lee:2022ymp}.
%
The values of $\beta_\ell$, $\beta_c$ were determined as described in the text.
}
\begin{center}
\begin{tabular*}
{\hsize}
{
l@{\extracolsep{0ptplus1fil}}
|l@{\extracolsep{0ptplus1fil}}
l@{\extracolsep{0ptplus1fil}}
l@{\extracolsep{0ptplus1fil}}
|l@{\extracolsep{0ptplus1fil}}}\hline\hline
        & $f\ $ & $ \alpha_0\ $ & $\alpha_1$ $[{\rm GeV}^{-2}]\ $ & $\beta$ $[{\rm GeV}^{-1}]\ $\\\hline
$\rho$ \cite{Cudell:1996sh, Pichowsky:1996tn} $\ $ & $\ell$ & $1.1\ $& $0.33\ $ & $2.80\ $ \\
$J/\psi$ \cite{Donnachie:1987pu, Lee:2022ymp} $\ $ & $ c \ $ & $1.25\ $ & $0.25\ $ & $0.076\ $\\
$J/\psi$ \mbox{\cite{ZEUS:2002wfj}}$\ $ & $ c \ $ & $1.2 \pm 0.009\ $ & $ 0.115 \pm 0.0118\ $ &
$0.11\ $
%
 \\[0.3ex]
 \hline\hline
\end{tabular*}
\end{center}
\end{table}

\section{Pomeron Parameters}
\label{SecPomParams}
With the transition loop integral calculated, the step to a vector meson photoproduction cross-section requires specification of the Pomeron parameters.
In the lighter quark sector, $\alpha_0$ was determined by fitting $pp$ and $\bar pp$ total cross-sections \cite{Cudell:1996sh}, with the value listed in Table~\ref{TabPom}.  Thereafter, $\alpha_1$ was fixed in Ref.\,\cite{Pichowsky:1996tn} by requiring a good description of the $t$ dependence of the $\pi N$ elastic scattering differential cross-section.
Using these parameters and our calculated form of the $\gamma\to \tfrac{1}{\surd 2}(u\bar u - d\bar d) + \mathbb P \to \rho^0$ transition loop, whose definition is obtained from Eq.\,\eqref{EqtqP} by straightforward replacements, we chose $\beta_\ell$ by requiring a good description of the differential cross-section for $\gamma p \to \rho^0 p$, with the value listed in Table~\ref{TabPom}.  The resulting comparison is drawn in Fig.\,\ref{Figrho0}.

Turning to the $c$ quark sector, Ref.\,\cite{Lee:2022ymp} analysed $J/\psi$ photoproduction by beginning with the original Pomeron parameters \cite{Donnachie:1987pu} and refitting selected $J/\psi$ photoproduction total cross-sections by tuning $\alpha_0$, with the result listed in Table~\ref{TabPom}.
That analysis used a VMD \emph{Ansatz} to replace the $\gamma \to \bar c c + {\mathbb P} \to J/\psi$ transition loop; so, the value of $\beta_c$ obtained in Ref.\,\cite{Lee:2022ymp} is inappropriate herein.
Using the Ref.\,\cite{Lee:2022ymp} values for $\alpha_{0,1}$ and $\beta_\ell$ from Table~\ref{TabPom}\,--\,Row~1, we introduced our result for the $\gamma \to \bar c c + {\mathbb P} \to J/\psi$ loop and determined $\beta_c$ by requiring a best fit to the $W\geq 26\,$GeV total cross-section data in Ref.\,\cite[H1\,2000]{H1:2000kis}.  Lower-$W$ data, available in other sources, were excluded in order to ensure that our analysis felt no influence from near-threshold information.  This procedure yielded the value of $\beta_c$ in Table~\ref{TabPom}\,--\,Row~2.

\begin{figure}[t]
\vspace*{0.5ex}

\leftline{\hspace*{0.5em}{\large{\textsf{A}}}}
\vspace*{-1ex}

\hspace*{1em}\includegraphics[width=0.45\textwidth]{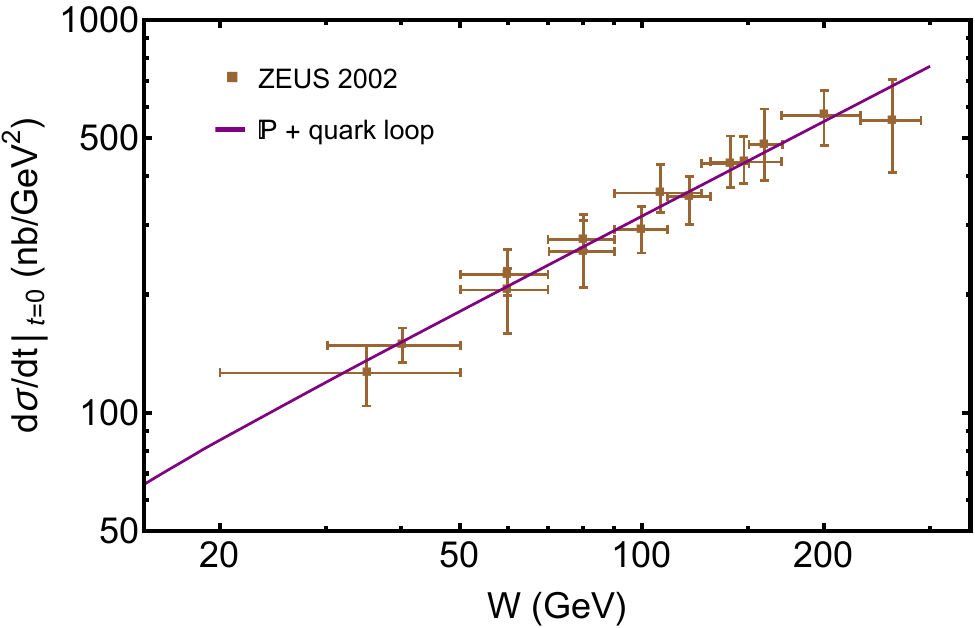}
\vspace*{0ex}

\leftline{\hspace*{0.5em}{\large{\textsf{B}}}}
\vspace*{-1ex}
\hspace*{1em}\includegraphics[width=0.45\textwidth]{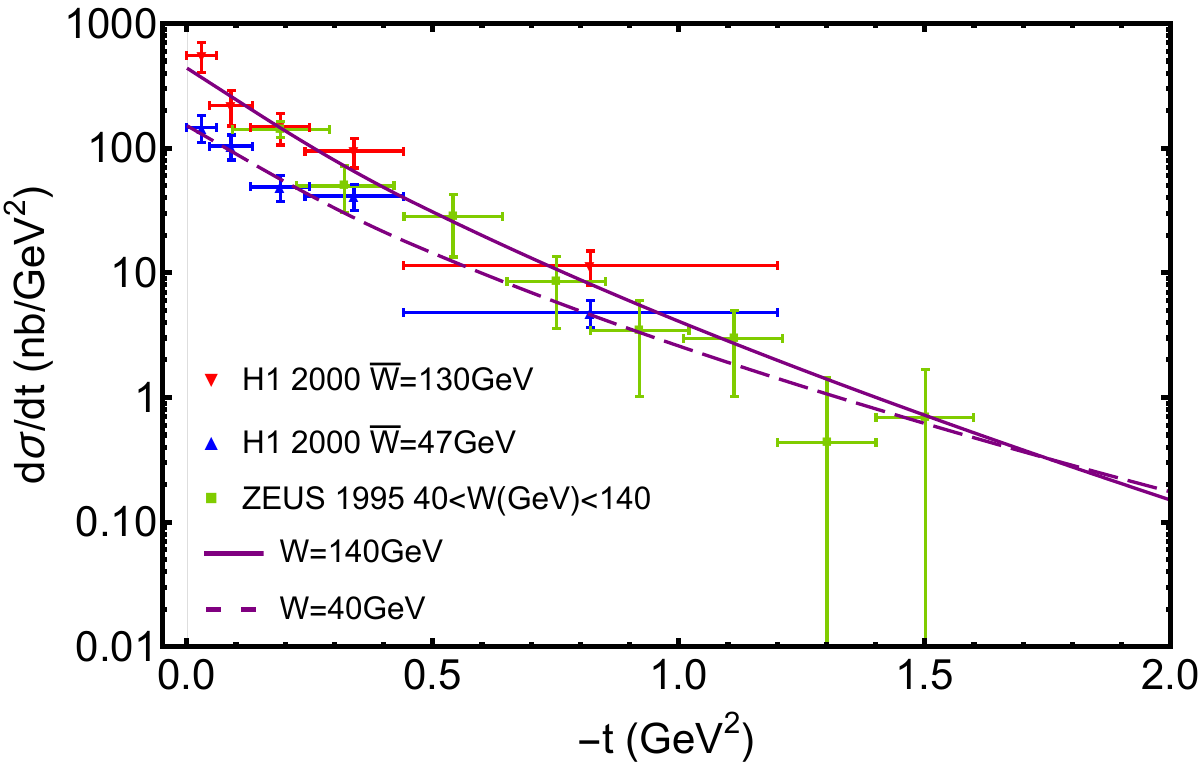}

\caption{\label{FigJpsi}
{\sf Panel A}.
Differential cross-section for $\gamma p \to  J/\psi p$ on $|t|/m_p\simeq 0$.
Solid purple curve: result obtained herein using the reaction model sketched in Fig.\,\ref{FigPomeron} and Pomeron parameters in Table~\ref{TabPom}\,--\,Row~3.
Data from Ref.\,\cite[ZEUS\,2002]{ZEUS:2002wfj}.
{\sf Panel B}.
Prediction herein for $t$-dependence of differential cross-section data from Ref.\,\cite[H1\,2000]{H1:2000kis} at two mean-$W$ values.  These data were not used to constrain the Pomeron parameters in Table~\ref{TabPom}\,--\,Row~3.
}
\end{figure}

More recent data on $\gamma p \to J/\psi p$, with $20< W/{\rm GeV} < 290$ is available in Ref.\,\cite[ZEUS\,2002]{ZEUS:2002wfj}, which also reports the differential cross-section on $|t|< 1.8\,$GeV$^2$.
Assuming validity of a Pomeron description, the associated parameters were determined therein by examining the $(W,t)$ dependence of the differential cross-section, with the results listed in Table~\ref{TabPom}\,--\,Row~3.
Employing these values to describe the Pomeron trajectory, we determined the associated value of $\beta_c$ by using our reaction model and requiring a least-squares best fit to the $W$-dependence of the $|t|\simeq 0$ cross-section, with the result listed in Table~\ref{TabPom}\,--\,Row~3.  The quality of the description is illustrated in Fig.\,\ref{FigJpsi}A.
It is worth noting that the Pomeron trajectory inferred in Ref.\,\cite[ZEUS\,2002]{ZEUS:2002wfj} is consistent with those extracted from data in Refs.\,\cite[H1\,2000]{H1:2000kis}, \cite[ZEUS\,2004]{ZEUS:2004yeh}: the error-weighted average of the three estimates yields $\alpha_0=1.20(1)$, $\alpha_1=0.112(11)$.
Importantly, as illustrated in Fig.\,\ref{FigJpsi}B, our reaction model also provides a good description of the $(W,t)$ dependence of the H1\,2000 differential cross-section.

With all Pomeron parameters determined, we are in a position to test our reaction model by delivering predictions for $\gamma p \to J/\psi p$ cross sections on the entire $W$ domain.

\begin{figure}[t]
\vspace*{0.5ex}

\leftline{\hspace*{0.5em}{\large{\textsf{A}}}}
\vspace*{-1ex}

\hspace*{1em}\includegraphics[width=0.45\textwidth]{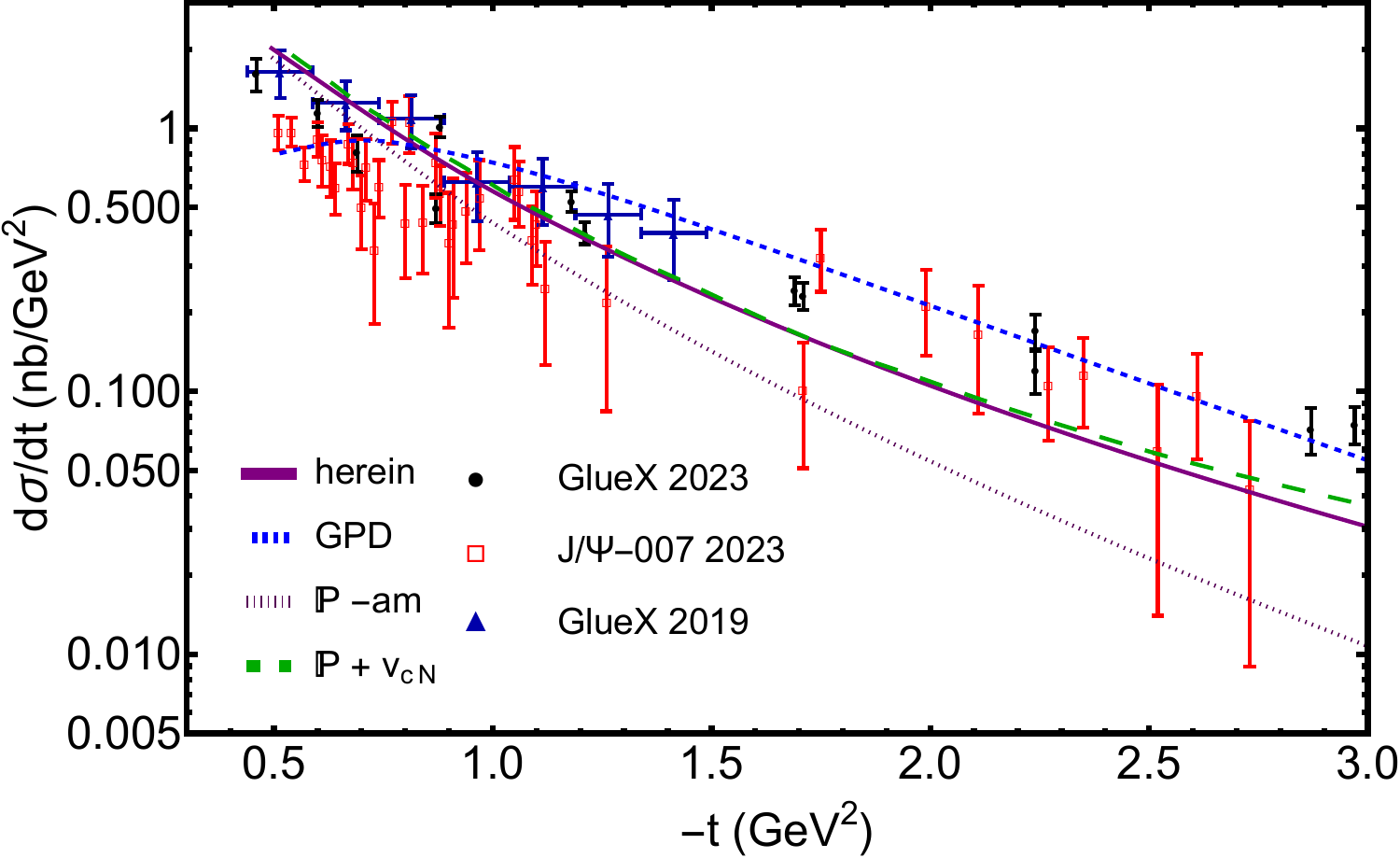}
\vspace*{0ex}

\leftline{\hspace*{0.5em}{\large{\textsf{B}}}}
\vspace*{-1ex}
\hspace*{1em}\includegraphics[width=0.45\textwidth]{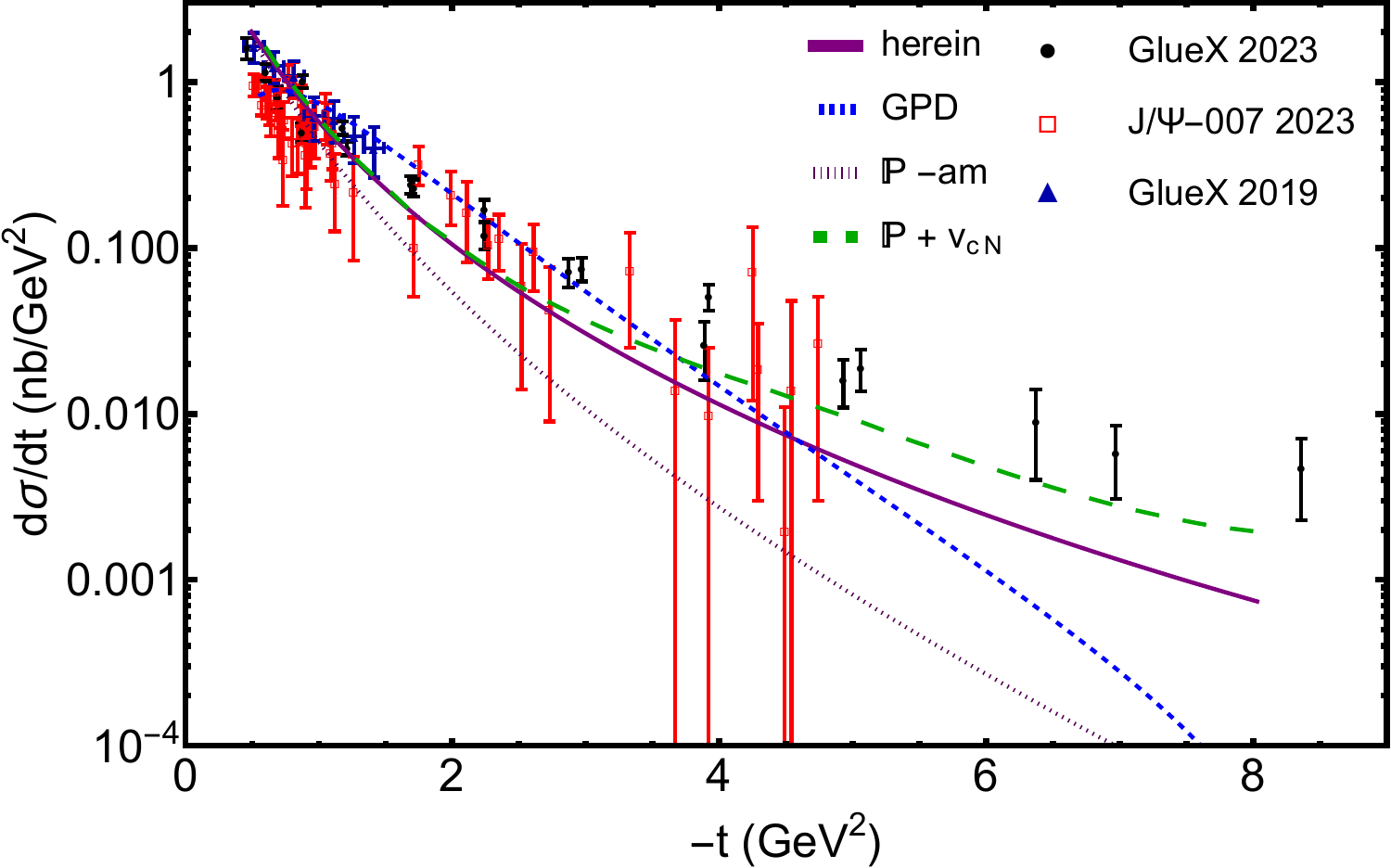}

\caption{\label{FigFinal0}
Near-threshold differential cross-section for $\gamma p \to  J/\psi p$.
Solid purple curve -- result obtained herein using the reaction model sketched in Fig.\,\ref{FigPomeron} and Pomeron parameters in Table~\ref{TabPom}\,--\,Row~3;
short-dashed blue curve -- GPD model \cite{Guo:2023pqw};
dashed green curve -- Pomeron + $J/\psi N$ scattering model \cite{Sakinah:2024cza};
and dotted purple curve $\mathbb P\,$-am.\ model.
%
Data from Refs.\,\cite[GlueX\,2019]{GlueX:2019mkq}, $4.43< W/{\rm GeV} < 4.8$;
\cite[GlueX\,2023]{GlueX:2023pev}, $4.38< W/{\rm GeV} < 4.63$;
\cite[$J/\psi$-007]{Duran:2022xag}, $4.41< W/{\rm GeV} < 4.54$.
(All theory curves computed using $W=4.5\,$GeV.)
}
\end{figure}

\begin{table}[t]
\caption{\label{TabX2}
$\chi^2/$dof for each curve drawn in the panels of Figure\,\ref{FigFinal0} --
herein, Ref.\,\cite[GPD]{Guo:2023pqw}, Ref.\,\cite[$\mathbb P + v_{cN}$]{Sakinah:2024cza}, and $\mathbb P$-am., computed in comparison with various combinations of available near-threshold data \cite[GlueX]{GlueX:2019mkq, GlueX:2023pev}, \cite[$J/\psi$-007]{Duran:2022xag}.
\emph{N.B}.\ The GPD model parameters were fitted to obtain a best fit to the differential cross-section data from these experiments; so, we have placed the results within quotation marks.
}
\begin{center}
\begin{tabular*}
{\hsize}
{
l@{\extracolsep{0ptplus1fil}}
|l@{\extracolsep{0ptplus1fil}}
l@{\extracolsep{0ptplus1fil}}
l@{\extracolsep{0ptplus1fil}}
|l@{\extracolsep{0ptplus1fil}}}\hline\hline
        & herein $\ $ & $\mathbb P + v_{cN}\ $ & $\mathbb P$-am.\ & GPD $\ $ \\\hline
GlueX $\ $ & $5.6\ $ & $5.2\ $ & $10.9\ $ & ``$6.0$'' $\ $ \\
$J/\psi$-007 $\ $ & $7.8 \ $ & $9.7\ $ & $\phantom{1}5.1\ $ & ``$2.6$'' $\ $\\
All $\ $ & $ 7.0 \ $ & $8.1\ $ & $ \phantom{1}7.2\ $ & ``$3.6$'' $\ $
%
 \\[0.3ex]
 \hline\hline
\end{tabular*}
\end{center}
\end{table}

\section{$J/\psi$ Differential Cross-Section Near Threshold}
\label{SecWT}
Our prediction for the $\gamma p \to J/\psi p$ differential cross section near threshold is drawn in Fig.\,\ref{FigFinal0} -- solid purple curve.
It is compared therein with data from Refs.\,\cite{GlueX:2019mkq, GlueX:2023pev, Duran:2022xag} and results from the reaction models in Refs.\,\cite{Guo:2023pqw, Sakinah:2024cza}.

Reference \cite{Guo:2023pqw} supposes, with some caveats, that the data is amenable to a GPD-based description, from which gluon gravitational form factors can be inferred.
%
In contrast, Ref.\,\cite{Sakinah:2024cza} argues that $J/\psi$ photoproduction is better described by Pomeron exchange, with the trajectory parameters in Table~\ref{TabPom}\,--\,Row~2, augmented by a $c$-quark+nucleon potential, $v_{cN}$, that drives $J/\psi \, N$ scattering and FSIs.  The FSIs dominate near threshold.
The final curve, $\mathbb P\,$-am., is obtained by beginning with the Ref.\,\cite{Sakinah:2024cza} model; setting $v_{cN}= 0$; replacing the Pomeron trajectory by that listed in Table~\ref{TabPom}\,--\,Row~3; and updating the Pomeron--$c$-quark coupling, using $\beta_c = 0.41\,$GeV$^{-1}$.
For each curve drawn in the panels of Fig.\,\ref{FigFinal0}, we list the $\chi^2/$degree-of-freedom (dof) in Table~\ref{TabX2}.

A first observation from Fig.\,\ref{FigFinal0} is that the momentum-dependence of the $\gamma\to c\bar c + \mathbb P \to J/\psi$ transition loop in Fig.\,\ref{FigPomeron} has a significant influence. 
This is highlighted by the difference between our prediction and the $\mathbb P\,$-am curve, in which the loop is expressed by a simple $t$-dependent $\mathbb P+J/\psi$ form factor \cite{Sakinah:2024cza}.

The GPD model \cite{Guo:2023pqw} assumes specific, simple forms for the two gluon gravitational form factors, and involves three parameters.  The parameters were fixed via a least-squares fit to the near threshold data \cite{GlueX:2019mkq, GlueX:2023pev, Duran:2022xag}.  As one may read ``by eye'' from Fig.\,\ref{FigFinal0} and is confirmed by Table~\ref{TabX2}, the description of $J/\psi$-007 data is better than that of the GlueX points.  This may be a signal of overfitting, the likelihood of which is amplified by the result for $d\sigma/dt$ being a concave function on the entire plotted domain, with a global maximum on $|t|\simeq 0.7\,$GeV$^2$.  All other reaction models deliver a differential cross-section that is a monotonically decreasing convex function.

Turning to the $\mathbb P + v_{cN}$ reaction model \cite{Sakinah:2024cza}, one sees a result in Fig.\,\ref{FigFinal0} that is similar to our prediction on $|t|\lesssim 2.5\,$GeV$^2$, albeit somewhat larger, and thereafter becomes harder.  Reviewing Table~\ref{TabX2} and comparing with the $\mathbb P$-am.\ column, it can be seen that the model's FSIs improve the description of GlueX data but degrade that of $J/\psi$-007.

Considering both Fig.\,\ref{FigFinal0} and Table~\ref{TabX2}, the most reasonable description of available near-threshold $d\sigma/dt$ data is provided by the reaction model presented herein: the parameter-free prediction delivers a convex differential cross-section and a $\chi^2/$dof that does not vary excessively across the data sets.

This discussion suggests that, whilst there are sound reasons to expect the proton mass radius, $r_m$, to be smaller than its charge radius, owing to the different character of the effective probe -- see Ref.\,\cite{Xu:2023bwv},  it is premature to draw any link between the Ref.\,\cite[$J/\psi$-007]{Duran:2022xag} data and a measurement of $r_m$ or, indeed, any other quantity inferred via a GPD-specific interpretation of those or other data.

\section{$J/\psi$ Total Cross-Section}
\label{SecWAll}
\begin{figure}[t]
\vspace*{1.2em}

\leftline{\hspace*{0.1em}{\large{\textsf{A}}}}
\vspace*{-2ex}
\includegraphics[width=0.46\textwidth]{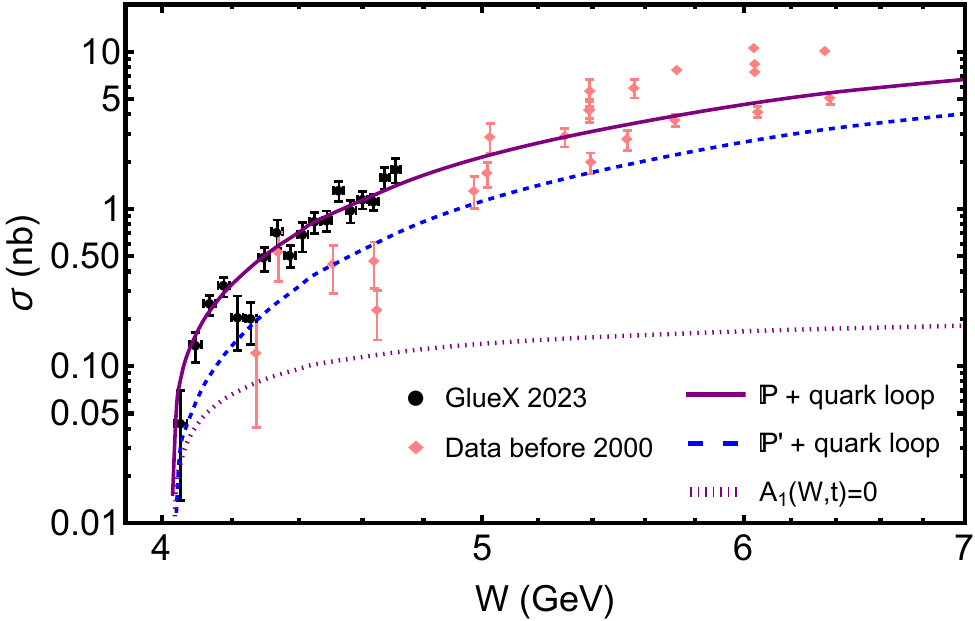}
\vspace*{0.1ex}

\leftline{\hspace*{0.1em}{\large{\textsf{B}}}}
\vspace*{-1ex}
\includegraphics[width=0.47\textwidth]{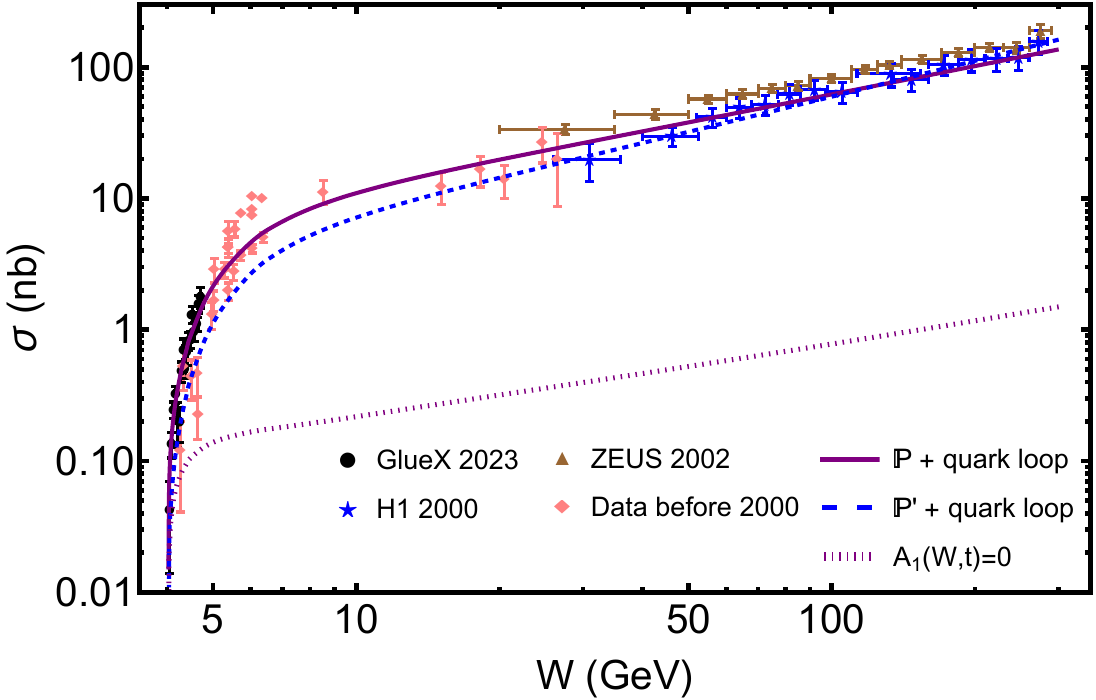}
\vspace*{-0.5ex}

\caption{\label{FigFinal1}
Total cross-section for $\gamma p \to  J/\psi p$.
Solid purple curve: result obtained herein using the reaction model sketched in Fig.\,\ref{FigPomeron} and Pomeron parameters in Table~\ref{TabPom}\,--\,Row~3.
Dashed blue: results obtained using Table~\ref{TabPom}\,--\,Row~2 parameters.
Dotted purple: result obtained when omitting the $A_1(W,t)$ component of the $\gamma\to c\bar c + \mathbb P \to J/\psi$ loop.
Data from Refs.\,\cite{GlueX:2019mkq, GlueX:2023pev, Camerini:1975cy, Gittelman:1975ix, Shambroom:1982qj, ZEUS:1995kab, H1:1996gwv, Amarian:1999pi,  H1:2000kis, ZEUS:2002wfj}.
}
\end{figure}

In Fig.\,\ref{FigFinal1}, we present our predictions for $\sigma_{\gamma p \to J/\psi p}$ as obtained using the reaction model sketched in Fig.\,\ref{FigPomeron} and the Pomeron parameters in Table~\ref{TabPom}\,--\,Row~3.
Panel A focuses on the near-threshold domain and \cite[GlueX]{GlueX:2019mkq, GlueX:2023pev} data; and
Panel B displays the entire $W$ range covered by data \cite{GlueX:2019mkq, GlueX:2023pev, Camerini:1975cy, Gittelman:1975ix, Shambroom:1982qj, ZEUS:1995kab, H1:1996gwv, H1:2000kis, ZEUS:2002wfj}.
Our reaction model provides a unified description of all available data on the entire $W$ domain.
Comparing the solid and dashed curves in Fig.\,\ref{FigFinal1}, the impact of the updated Pomeron trajectory is plain. 
Moreover, the difference between both and the dotted curves in Fig.\,\ref{FigFinal1}, obtained by setting $A_1(W,t) \equiv 0$ in $t_{\mu\alpha\nu}$, serves to illustrate the overwhelming dominance of the $A_1(W,t)$ component of the $\gamma\to c\bar c + \mathbb P \to J/\psi$ loop.

\begin{figure}[t]
\vspace*{1.2em}

\leftline{\hspace*{0.1em}{\large{\textsf{A}}}}
\vspace*{-2ex}
\includegraphics[width=0.46\textwidth]{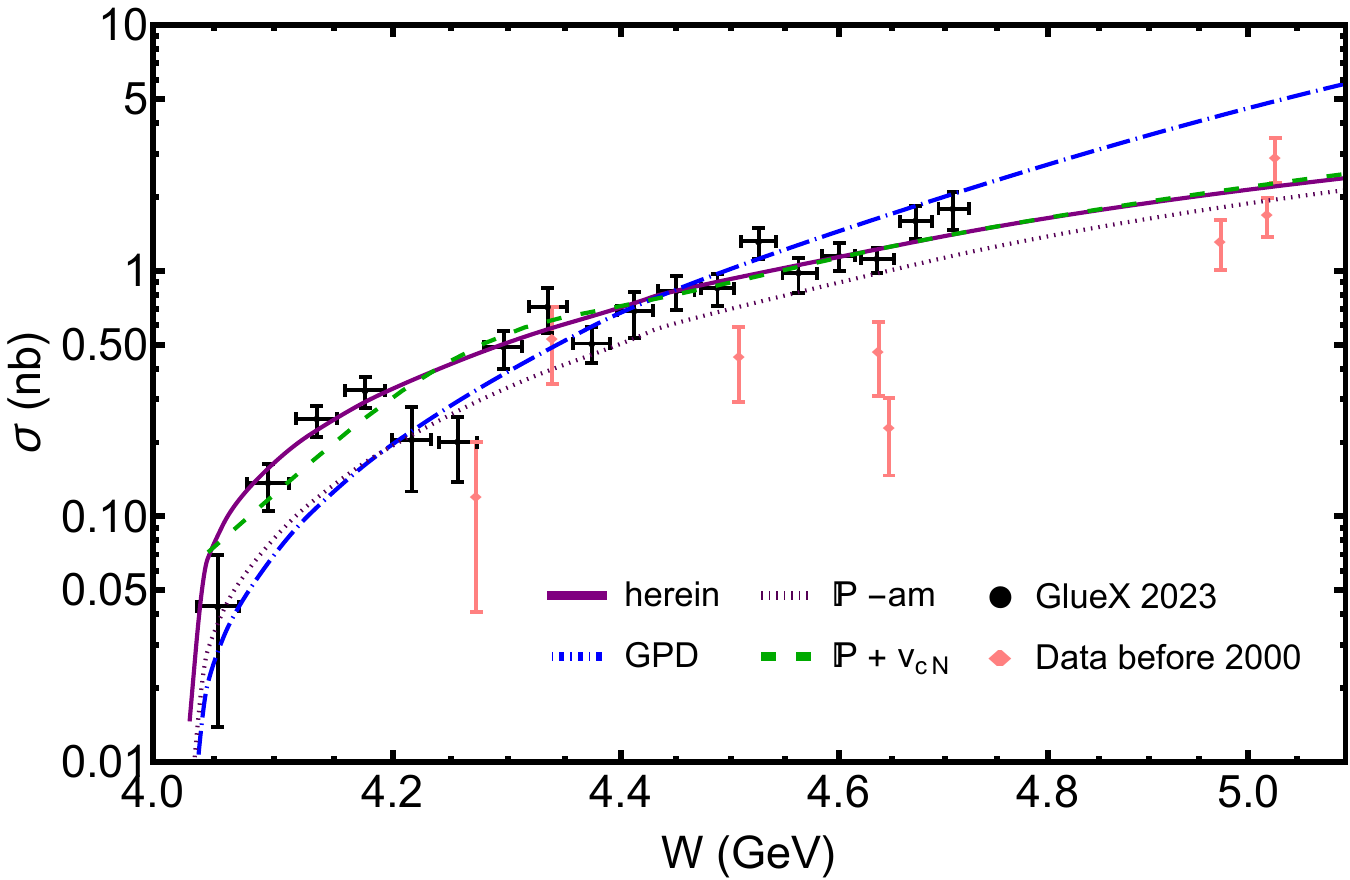}
\vspace*{0.1ex}

\leftline{\hspace*{0.1em}{\large{\textsf{B}}}}
\vspace*{-1ex}
\includegraphics[width=0.46\textwidth]{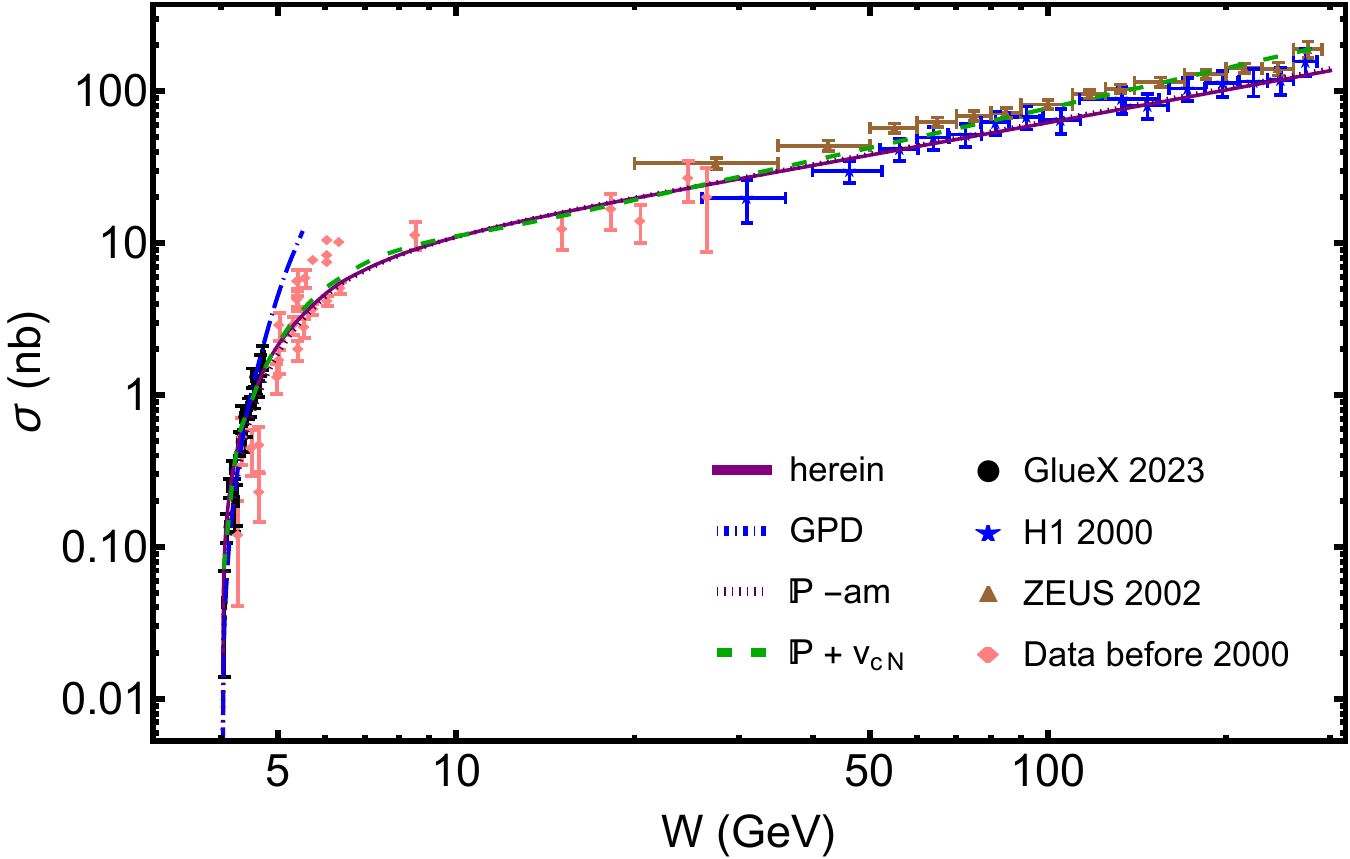}
\vspace*{-0.5ex}

\caption{\label{FigFinal2}
Total cross-section for $\gamma p \to  J/\psi p$.
Solid purple curve: result obtained herein using the reaction model sketched in Fig.\,\ref{FigPomeron} and Pomeron parameters in Table~\ref{TabPom}\,--\,Row~3.
Dot-dashed blue curve \cite{Guo:2023pqw}: GPD model, which is undefined away from threshold.
Dashed green curve \cite{Sakinah:2024cza}: $J/\psi N$ final-state interactions (FSIs) plus Pomeron, described by trajectory in Table~\ref{TabPom}\,--\,Row~2.
Dotted purple curve: amended version of the Ref.\,\cite{Sakinah:2024cza} reaction model, eliminating FSI component and using the Table~\ref{TabPom}\,--\,Row~3 Pomeron trajectory.
Data from Refs.\,\cite{GlueX:2019mkq, GlueX:2023pev, Camerini:1975cy, Gittelman:1975ix, Shambroom:1982qj, ZEUS:1995kab, H1:1996gwv, Amarian:1999pi, H1:2000kis, ZEUS:2002wfj}.
}
\end{figure}

In Fig.\,\ref{FigFinal2} we compare our prediction with those obtained using alternative reaction models, \emph{viz}.\ \cite[GPD]{Guo:2023pqw}, \cite[$\mathbb P +v_{cN}$]{Sakinah:2024cza}, and $\mathbb P$-am.
The $\chi^2/$dof for each curve is listed in Table~\ref{TabX2B}.


\begin{table}[b]
\caption{\label{TabX2B}
$\chi^2/$dof associated with selected reaction models when compared with data on different $W$ (in GeV) ranges.
Row~1: $W<5\,$ -- \cite[GlueX]{GlueX:2019mkq, GlueX:2023pev}.
Row~2: $W>20$ -- \cite[H1, ZEUS]{H1:2000kis, ZEUS:2002wfj}.
Row~3: All $W$ -- \cite[GlueX]{GlueX:2019mkq, GlueX:2023pev} and \cite[H1, ZEUS]{H1:2000kis, ZEUS:2002wfj}.
Row~4 -- $W>20$ -- \cite[H1]{H1:2000kis} only.
Row~5: All $W$ -- \cite[GlueX]{GlueX:2019mkq, GlueX:2023pev} and \cite[H1]{H1:2000kis} only.
The entry ``undefined'' highlights that the model cannot be applied beyond the near-threshold region.
Ref.\,\cite[ZEUS]{ZEUS:2002wfj} reports a large uncertainty in $W$, but a small uncertainty in $\sigma$.  This distorts the $\chi^2$; so, the list also contains entries that exclude ZEUS from the $\chi^2$ evaluation.
%
%
}
\begin{center}
\begin{tabular*}
{\hsize}
{
l@{\extracolsep{0ptplus1fil}}
|l@{\extracolsep{0ptplus1fil}}
|l@{\extracolsep{0ptplus1fil}}
l@{\extracolsep{0ptplus1fil}}
|l@{\extracolsep{0ptplus1fil}}}\hline\hline
(GeV) & herein  $\ $ & $\mathbb P + v_{cN}$ \cite{Sakinah:2024cza} $\ $ & $\mathbb P\,$-am.\  $\ $ &
GPD \cite{Guo:2023pqw} $\ $  \\\hline
$W<5\ $ & $1.4\ $  & $1.7\ $ & $3.0\ $ & $2.5\ $  \\
$W> 20 $ & $5.7\ $  & $2.8\ $ & $5.3\ $ & undefined \\
All $W$  & $3.5\ $ & $2.1\ $ & $4.1\ $ & undefined  \\
$W> 20\ _{\rm H1}\ $ & $0.42\ $& $2.1\ $ & $0.40\ $ & undefined  \\
All $W_{\rm H1}$ & $1.0\ $ & $1.6\ $ & $2.1\ $ & undefined
 \\[0.2ex]
 \hline\hline
\end{tabular*}
\end{center}
\end{table}


Regarding the GPD model, the best-fit curve from Ref.\,\cite{Guo:2023pqw} is drawn in Fig.\,\ref{FigFinal2}: in comparison with \cite[GlueX]{GlueX:2019mkq, GlueX:2023pev} total cross-section data, the reduced-$\chi^2$ is fair -- see Table~\ref{TabX2B}.  However, the model cannot meet the requirement of a unified description of all data.  Moreover, even on the subdomain of assumed applicability, the description of \cite[GlueX]{GlueX:2019mkq, GlueX:2023pev} data does not match that produced by the reaction model proposed herein.  These facts further undermine any attempt to connect near threshold $J/\psi$ photoproduction data with the proton's gluon GPD.



Considering the result produced by the reaction model in Ref.\,\cite{Sakinah:2024cza},
one reads from Fig.\,\ref{FigFinal2} and Table~\ref{TabX2B} that this approach can deliver a good description of available total cross-section data on the entire $W$ range.
In this case, regarding also the description of differential cross-section data -- see Fig.\,\ref{FigFinal0} and Table~\ref{TabX2}, it is the link drawn between such agreement and dominance of FSIs near threshold that is uncertain.
The connection relies on a particular choice for the Pomeron trajectory, \emph{i.e}., Table~\ref{TabPom}\,--\,Row~2.
If one instead uses the trajectory from Table~\ref{TabPom}\,--\,Row~3 and $\beta_c = 0.41\,$GeV$^{-1}$, \emph{i.e}., the $\mathbb P$-am.\ variant, then a practically equivalent description of data is possible without calling upon FSIs -- see the dotted purple curves in Fig.\,\ref{FigFinal2}.

The reaction model we have elucidated -- sketched in Fig.\,\ref{FigPomeron} -- combines the parameter-free CSM prediction for the momentum dependence of the $\gamma\to c\bar c + \mathbb P \to J/\psi$ transition loop with Pomeron exchange described by the trajectory determined in Ref.\,\cite[ZEUS\,2002]{ZEUS:2002wfj}, which is practically indistinguishable from that inferred in Refs.\,\cite[H1\,2000]{H1:2000kis}, \cite[ZEUS\,2004]{ZEUS:2004yeh}.  Here, eliminating VMD assumptions, novel effects of photon + quark + meson dynamics are expressed by revealing the quark loop involved in the photoproduction process.
Looking at Figs.\,\ref{FigFinal0} -- \ref{FigFinal2} and Tables~\ref{TabX2}, \ref{TabX2B}, our reaction model currently provides the best description of all available data; both that newly obtained near threshold (differential and total cross-sections) and also the long existing data reaching out to $W \approx 300\,$GeV.

\section{Summary and Perspective}
We have described a reaction model for $\gamma + p \to J/\psi + p$ photoproduction that exposes the $c \bar c$ content of the dressed photon in making the transition $\gamma\to c\bar c + \mathbb P \to J/\psi$ and couples the intermediate $c \bar c$ system to the proton's valence quarks via Pomeron exchange [Fig.\,\ref{FigPomeron}].  The only parameters in this model are those characterising the Pomeron trajectory (two) and the Pomeron+$c$-quark coupling (one).  With those parameters fitted to high-$W$ data reported in Ref.\,\cite[ZEUS\,2002]{ZEUS:2002wfj} [Sect.\,\ref{SecPomParams}], our reaction model provides a uniformly good description of differential and total cross-sections for $J/\psi$ photoproduction from the proton on the entire kinematic range measured to date. 
[Sects.\,\ref{SecWT}, \ref{SecWAll}].  The quality of the description is typically better than that provided by alternative models.

Our reaction model is viable, being sufficient to describe existing data.  However, we do not pretend that it is necessarily complete.  Most importantly, we judge this analysis to be valuable in highlighting that it is premature to interpret existing $\gamma + p \to J/\psi + p$ data in terms of in-proton gluon distributions, as a window onto the QCD trace anomaly, or as providing insights into pentaquark production.
The desire to make such interpretations should be tempered whilst development of reaction models continues and higher precision data are accumulated.
Then, subsequent mutual feedback could lead to a tightly constrained phenomenology that may be used in understanding the $\gamma + p \to J/\psi + p$ process and building a solid bridge between the associated data and proton properties.


%
\medskip
\noindent\textbf{Acknowledgments}.
We are grateful to Y.~Guo, T.-S.\ H.~Lee, V.~Mokeev, I.~Strakovsky and J.-J.~Wu for valuable discussions; and are particularly indebted to T.-S.\ H.~Lee for providing the Ref.\,\cite{Sakinah:2024cza} reaction model results over the entire $W$ range, as displayed in Fig.\,\ref{FigFinal2}B, and to J.-J.~Wu for assisting us in comparing our predictions with related results in Ref.\,\cite{Wu:2012wta}.
Work supported by:
National Natural Science Foundation of China (grant nos.\,12135007, 12233002);
and
Natural Science Foundation of Jiangsu Province (grant no.\ BK20220122).

\medskip
\noindent\textbf{Data Availability Statement}. This manuscript has no associated data or the data will not be deposited. [Authors' comment: All information necessary to reproduce the results described herein is contained in the material presented above.]

\medskip
\noindent\textbf{Declaration of Competing Interest}.
The authors declare that they have no known competing financial interests or personal relationships that could have appeared to influence the work reported in this paper.


\end{document}